\documentclass{aastex6}

\fullcollaborationName{The Friends of AASTeX Collaboration}
\begin{document}

\title{Formation of an active region filament driven by a series of jets}

\author{Jincheng Wang\altaffilmark{1,2,3}, Xiaoli Yan\altaffilmark{1,3}, Zhongquan Qu\altaffilmark{1,2,3}, Satoru UeNo\altaffilmark{4}, Kiyoshi Ichimoto\altaffilmark{4}, Linhua Deng\altaffilmark{1,3,5}, Wenda Cao\altaffilmark{6}, Zhong Liu\altaffilmark{1,3}}

\altaffiltext{1}{Yunnan Observatories, Chinese Academy of Sciences, Kunming 650011, P. R. China.}
\altaffiltext{2}{University of Chinese Academy of Sciences, Yuquan Road, Shijingshan Block Beijing 100049, P. R. China.}
\altaffiltext{3}{Center for Astronomical Mega-Science, Chinese Academy of Sciences, 20A Datun Road, Chaoyang District, Beijing, 100012, P. R. China}
\altaffiltext{4}{Hida Observatory, Kyoto University, Kamitakara, Takayama, Gifu 506-1314, Japan}
\altaffiltext{5}{School of Software Engineering, Chongqing University of Arts and Sciences, Chongqing 402160, P. R. China}
\altaffiltext{6}{Big Bear Solar Observatory, 40386 North Shore Lane, Big Bear City, CA 92314, USA}
\begin{abstract}
We present a formation process of a filament in active region NOAA 12574 during the period from 2016 August 11 to 12. Combining the observations of GONG H$\alpha$, Hida spectrum and SDO/AIA 304 \rm\AA, the formation process of the filament is studied. It is found that cool material ($T\sim10^4$ K) is ejected by a series of jets originating from the western foot-point of the filament. Simultaneously, the magnetic flux emerged from the photosphere in the vicinity of the western foot-point of the filament. These observations suggest that cool material in the low atmosphere can be directly injected into the upper atmosphere and the jets are triggered by the magnetic reconnection between pre-existing magnetic fields and new emerging magnetic fields. Detailed study of a jet at 18:02 UT on August 11 with GST/BBSO TiO observations reveals that some dark threads appeared in the vicinity of the western foot-point after the jet and the projection velocity of plasma along the filament axis was about 162.6$\pm$5.4 km/s. Using with DST/Hida observations, we find that the injected plasma by a jet at 00:42 UT on August 12 was rotating. Therefore, we conclude that the jets not only supplied the material for the filament, but also injected the helicity into the filament simultaneously. Comparing the quantity of mass injection by the jets with the mass of the filament, we conclude that the estimated mass loading by the jets is sufficient to account for the mass in the filament.
\end{abstract}

\keywords{Sun:filaments, Sun:evolution, Sun:activity}

\section{Introduction} \label{sec:intro}
Solar filaments/prominences are one of the most common structures in the corona, which may lead to energetic coronal mass ejections (CMEs) and flares when they erupt \citep{che00,for00,lin00,yan13,xue16,yang17}. They appear at the limb as bright features called prominences. In contrast, they appear on the disk as dark filamentary structures called filaments. It is widely accepted that solar filaments are cool, dense plasma structures suspended in the extremely hot solar corona. They often lie above magnetic polarity inversion lines (PILs) on the photosphere and are supported by the local magnetic fields (e.g. magnetic dips or twisted structures) against the gravity \citep{bab55,mar98,yang14}. Generally, according to their locations on the solar disk, filaments can be classified as active region filaments, intermediate filaments, and quiescent filaments \citep{pat02,mac10}. Observationally, active region filaments are lower, smaller and shorter-lived than quiescent or intermediate filaments. However, active region filaments are more likely to erupt than other two types \citep{jin04,par14}.

How cool and dense material is supplied to filaments embedded in the hot and tenuous corona remains an open question. On the one hand, many researchers investigated the formation of filament magnetic field structures and proposed two different efficient ways to form the magnetic field structures of filaments: surface effect \citep{van89,mar01,yan15,yan16,yang16,wan17,xue17,chen18}  and subsurface effect \citep{oka08,oka09,lit09,lit10,mact10,yan17}. \cite{mact10} analyzed a set of MHD simulations on filament formation and supported the observation reported by \cite{oka08,oka09} that there was a flux rope emerging under the filament. However, \cite{var12} reanalyzed the data and found that the observational evidence was not enough to support the emergence of a flux rope. It has been reported that the flux cancellation and convergence in the photosphere play an important role in the formation of filament magnetic structures \citep{wan01,wan07}.

Material origination of the filament is another issue for understanding the formation of the filament. It is believed that the material of the large filament must come from the low solar atmosphere (chromosphere), because not enough plasma could be supplied for these large filaments in the corona \citep{pik71,zir94}. Comparing the elemental abundances in situ observations with that of the photosphere, \cite{son17} suggested that the plasma of an active region filament may originate from the lower solar photosphere instead of the corona.

The mechanisms by which low atmospheric material is converted into filament material are not fully understood. According to numerous observations and numerical simulations, three popular models have been proposed by many authors: injection model, levitation model, and evaporation-condensation model \citep{mac10}. The injection model suggests that the cool plasma is forced upward into the filament channels or typical filament heights through sufficient magnetic forces (e.g. magnetic reconnection in the vicinity of PILs or near the filament foot-points) \citep{pri96,lit99,wan99,wan01}. \cite{cha03} reported that the plasma could be ejected into an active-region filament by a successive of jets and small eruptions through magnetic reconnection in the vicinity of the magnetic PIL. \cite{liu05} reported that the formation of two new filaments is closely correlated with surges and the cool material is directly injected into the main axis/channels of filaments by these surges at one foot-point. \cite{zou16} proposed that the cool material was injected in the form of fibrils to replenish the filament and proposed that the magnetic reconnection play an important role in transporting cool material into the filament.

The levitation model proposes that cool plasma is directly lifted by rising magnetic fields at the magnetic PIL. In one scenario of this model, the highly twisted flux rope emerges from the photosphere and brings up cool plasma in the axis of rising twisted flux rope \citep{rus94,gal99,lit05,kuc12}.

Based on the fact that adding heat to a coronal loop increases the density of the corona accompanied by decreasing slightly the chromospheric mass, the evaporation-condensation model has been proposed. According to numerous numerical simulations, many researchers found that the cool material could be condensed at the apex of coronal loop by heating near the foot-points \citep{ant91,kar05,kar06,xia11,xia12,kan17}. Using the observations from SDO/AIA, \cite{liu12} considered that the coronal condensation should be responsible for the formation of a prominence. Although these models can explain some observations of the filament formation, the physical mechanisms of material injection are still hardly to understand fully. Besides, due to the long period of formation process, the entire formation process of filaments, especially quiescent filaments, is difficult to be captured.

Compared with quiescent filaments and intermediate filaments, active region filaments need relatively short time to form. It is about several hours or one day for forming an active region filament in general. In this paper, we investigate a formation process of a filament in active region NOAA 12574 during the period from 2016 August 11 to 12. The whole formation process that the filament was from absent to present can be exhibited clearly, which is a rare case for understanding the formation of the filament. By the ground-based and space-based observations, the material injection of the filament is studied in detail. We investigate jetting events as possible mechanisms for loading mass into the active region filament. We compare the cumulative mass flux supplied by the jets with the estimated mass in the filament. Furthermore, we estimate the magnetic windings injected into the filament system.

\section{Observations and Methods} \label{sec:obser methods}
During the period from 2016 August 11 to 12, a filament formed gradually in active region NOAA 12574, which was located in the northeast hemisphere (eg.about (-375$\arcsec$, 50$\arcsec$)). Observations from the Global Oscillation Network Group \citep{har96,har11} and the \emph{Solar Dynamics Observatory} \citep{pes12} covered the whole process of the filament formation. Combined with the observations by $Hinode$ \citep{kos07}, Goode Solar Telescope \citep{cao10}, Domeless Solar Telescope \citep{nak85} and \emph{Interface Region Imaging Spectrograph} \citep{pon14}, two material injection events (two jets) related to the formation of the filament are studied in detail.

Full disk H$\alpha$ images from six GONG stations are used to monitor filament formation and evolution over two days. The CCD plate-scale of the H$\alpha$ images is about 1$\arcsec$ $\rm pixel^{-1}$ and the cadence is 1 minutes. The data from Atmospheric Imaging Assembly \citep{lem12} and the Helioseismic and Magnetic Imager \citep{scher12,schou12} on board the \emph{SDO} provides full-disk, multi-wavelength, high spatio-temporal resolution observations for this study. The SDO/AIA has seven extreme ultraviolet (EUV) and three ultraviolet-to-visible (UV) channel images with the CCD plate-scale of 0$\farcs6$ $\rm pixel^{-1}$. The cadence of EUV and UV channel images are 12 s and 24 s, respectively. Using the 6173 $\rm \AA$ Fe $\rm I$ absorption line, the SDO/HMI can provide the Doppler shift, line-of-sight magnetic field, continuum intensity and vector magnetic field on the solar photosphere with the CCD plate-scale of 0$\farcs$5 $\rm pixel^{-1}$, with the cadences of the three former channels being about 45s and that of the latter one being about 12 minutes. The SDO/AIA 304 $\rm \AA$ images exhibit the formation process of the active region filament, while the SDO/AIA 1600 $\rm \AA$ images and SDO/HMI line-of-sight magnetic fields are utilized to investigate the physical mechanisms of the material injection related to the filament formation.

TiO (7057 $\rm\AA$) images observed by the Broadband Filter Imagers on the 1.6 m GST at the \emph{Big Bear Solar Observatory} with the CCD plate-scale of 0$\farcs$0342 $\rm pixel^{-1}$ and a cadence of 15 seconds provide important information in the photosphere during a jet at 18:02 UT on August 11. The Spectro-polarimeter instrument \citep{ich08} of the Solar Optical Telescope \citep{tsu08} on board $Hinode$ provides the photospheric vector magnetic field and Doppler shift with the CCD plate-scale of 0$\farcs$16 $\rm pixel^{-1}$  for this study. The vector magnetic fields are derived by performing the Milne-Eddington (M-E) inversion of the spectro-polarimetric profiles of two magnetically Fe lines at 6301.5 $\rm\AA$ and 6302.5 $\rm\AA$ \citep{oro07}. The retrieval of the vector magnetic field is carried out by the data analysis pipeline at HAO/CSAC (Community Spectro-polarimetric Analysis Center of High Altitude Observatory, Boulder). The data are referred to as \emph{Hinode} Level 2 data sets at HAO/CSAC.

Spatially scanned H$\alpha$ and Ca II K spectrums with the  Horizontal Spectrograph (HS) obtained simultaneously by DST at Hida Observatory \citep{uen04} are used to investigate a material injection event (Jet B) at 00:42 UT on August 12. The spectral sampling is 0.020 $\rm \AA$ and the angular sampling along the slit is 0$\farcs$24. The full wavelength-range of the obtained spectrum is 16 $\rm \AA$ and it scans the observational region at every 15 s. The spatial scan step is about 0$\farcs$64 and the time interval between two spatial steps is 0.05 s. Based on the H$\alpha$ spectral, we can calculate the Doppler velocity of H$\alpha$ by the follow equation: $v_{dop}=((\lambda_{obs}-\lambda_0)/\lambda_0)*c$, where $\lambda_{obs}$ is the center of observational line, $\lambda_{0}$ is the center of H$\alpha$ line and $c$ is the velocity of light. We use the weight-reverse-intensity method to derive the $\lambda_{obs}$ of H$\alpha$ line \citep{su16}. These velocities represent the ``mean line-of-sight velocity" of the material along the light path, which are justified for estimating direction of the motion (blue shift or red shift) in this study. These velocities are inferred by assuming that the mean velocity in the quiet region is zero. IRIS can provide spectral scan and slit-jaw images (SJIs) at near-ultraviolet (NUV) and far-ultraviolet (FUV) lines simultaneously \citep{pon14}. The 1400 $\rm \AA$ SJI and Si $\rm {IV}$ spectrum of IRIS are used to investigate the jet in this study. The single Gaussian fitting method is used to obtain the center of observational Si $\rm {IV}$ line for calculating the Doppler velocity of Si $\rm {IV}$ line, which allows us to investigate the properties of the jet nearby the one foot-point of the filament.

\begin{figure}[ht!]
\figurenum{1}
\plotone{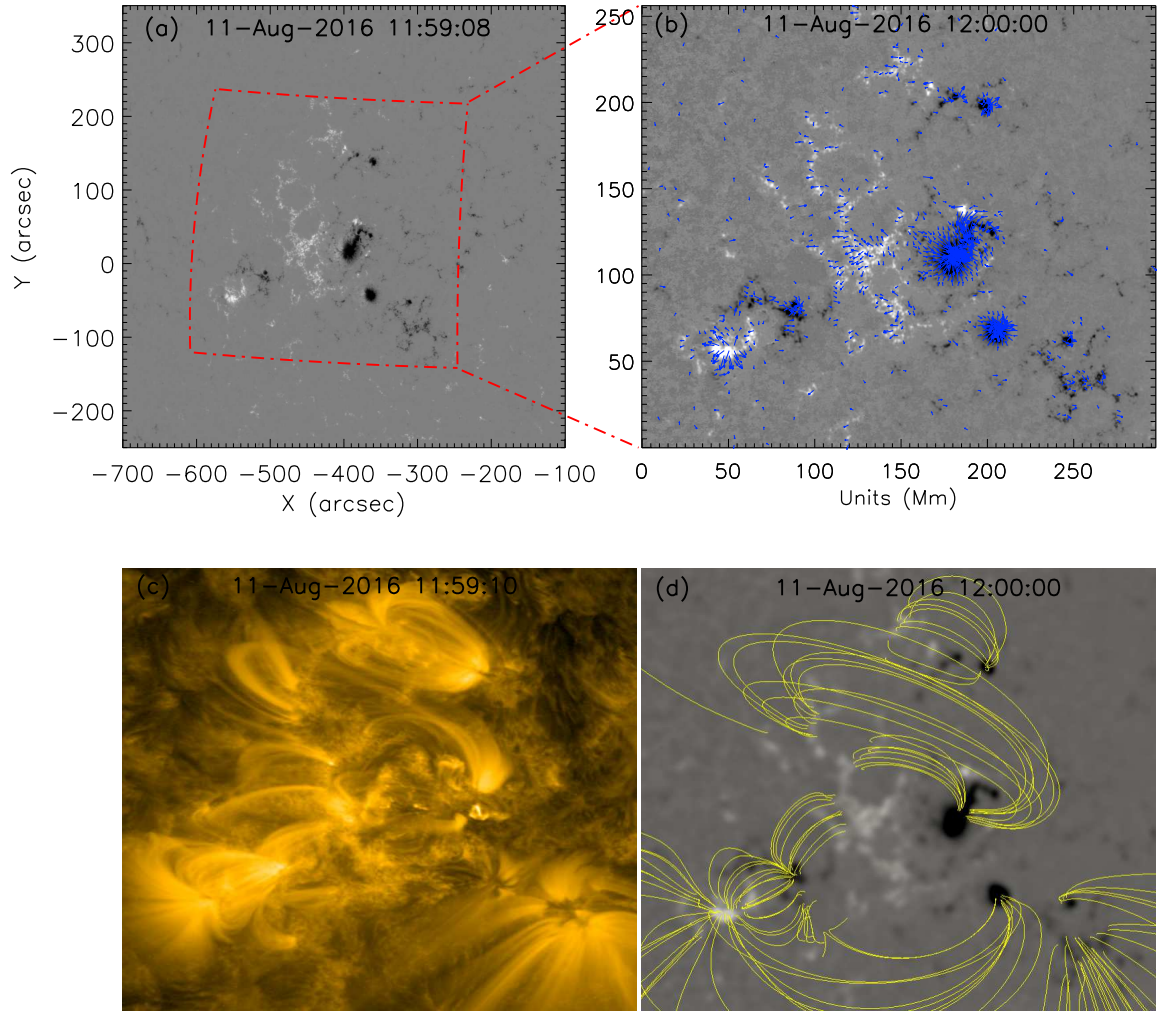}
\caption{Observations of active region NOAA 12574 at about 12:00 UT on 2016 August 11. (a) Line-of-sight magnetic field from SDO/HMI. White denotes the magnetic field with positive polarity, while black denotes the magnetic field with negative polarity. (b) Photospheric vector magnetogram from SDO/HMI, corresponding to the region marked by the red dotted-dashed box in panel (a). The blue arrows indicate the transverse field. (c) SDO/AIA 171 $\rm\AA$ image. (d) The magnetic field lines derived from the NLFFF model extrapolation. \label{figure0}}
\end{figure}

In order to illustrate the morphological structure of the magnetic fields of the jets and filament, we reconstruct the coronal magnetic field through a nonlinear force-free field (NLFFF) model with vector magnetic field observed by SDO/HMI. NLFFF extrapolation is obtained by using the ``weighted optimization" method \citep{whe00,wie04} after preprocessing the photospheric boundary to meet the force-free condition \citep{wie06}. Before the extrapolation, we use a 2 $\times$ 2 rebinning of the boundary data to 0.72 Mm pixel$^{-1}$ as some authors did \citep{sun12,wie12,liu13}. The extrapolated field derived by the NLFFF model extrapolation is thought to well match the magnetic structures of the observational images \citep{wie05}. From panels (c) and (d) of the Fig.\ref{figure0}, the magnetic field lines produced by the NLFFF model extrapolation are roughly compatible with the EUV loops observed by SDO/AIA 171 \AA. On the other hand, the magnetic helicity injection rate across a surface S can be calculated by using the following equation \citep{ber84}:
\begin{equation}\label{equ3}
  \frac{dH}{dt}=-2\int_S(\textbf{A}\cdot \textbf{u})B_n dS,
\end{equation}
in which A is the vector potential of the potential field, $\textbf{u}$ denotes the velocity of the flux tubes on the boundary (the flux transport velocity, $\textbf{u}=\textbf{V}_t - (V_n/B_n)\textbf{B}_t$) \citep{dem03}, and $B_n$ denotes the strength of normal component of the magnetic field. With a acceptable assumption that the solar photosphere S is planar, \citet{par05} proposed the helicity injection rate could be transformed to:
\begin{equation}\label{equ4}
  \frac{dH}{dt}=-\frac{1}{2\pi}\int_{S\arcmin} \int_S \frac{d\theta(\bf{r})}{dt}B_nB\arcmin_ndSdS\arcmin,
\end{equation}
and
\begin{equation}\label{equ5}
  \frac{d\theta(\bf{r})}{dt}=\frac{1}{r^2}(\bf{r} \it \times \frac{dr}{dt})_n=\frac{1}{r^2}(\bf r \times (u-u\arcmin))\it _n,
\end{equation}
 where $ \bf r = x-x\arcmin$ denotes the vector between two photospheric points defined by $x$ and $x\arcmin$, $\textbf{u}$ and $\textbf{u\arcmin}$ are the homologous velocities of two different points. A good proxy of helicity flux density $G_\theta(x)$ can be then defined as:
\begin{equation}\label{equ6}
  G_\theta(x)=-\frac{B_n}{2\pi}\int_{S\arcmin}
  \frac{d\theta(\bf{r})}{dt}B\arcmin_ndS\arcmin.
\end{equation}
 We use the differential affine velocity estimator for vector magnetograms (DAVE4VM) method to compute the velocity of the flux tubes \citep{sch08}. The whole active region of the vector magnetograms (see Fig.\ref{figure0} (b)) are used to compute the helicity flux density. The method is similar to that used by \cite{jin12}. Once the helicity flux density has been determined, the helicity flux can be integrated by the integral region using the equation (\ref{equ4}).

\section{Results} \label{sec:results}
\subsection{The formation process of the filament} \label{subsec:process}
\begin{figure}[ht!]
\figurenum{2}
\plotone{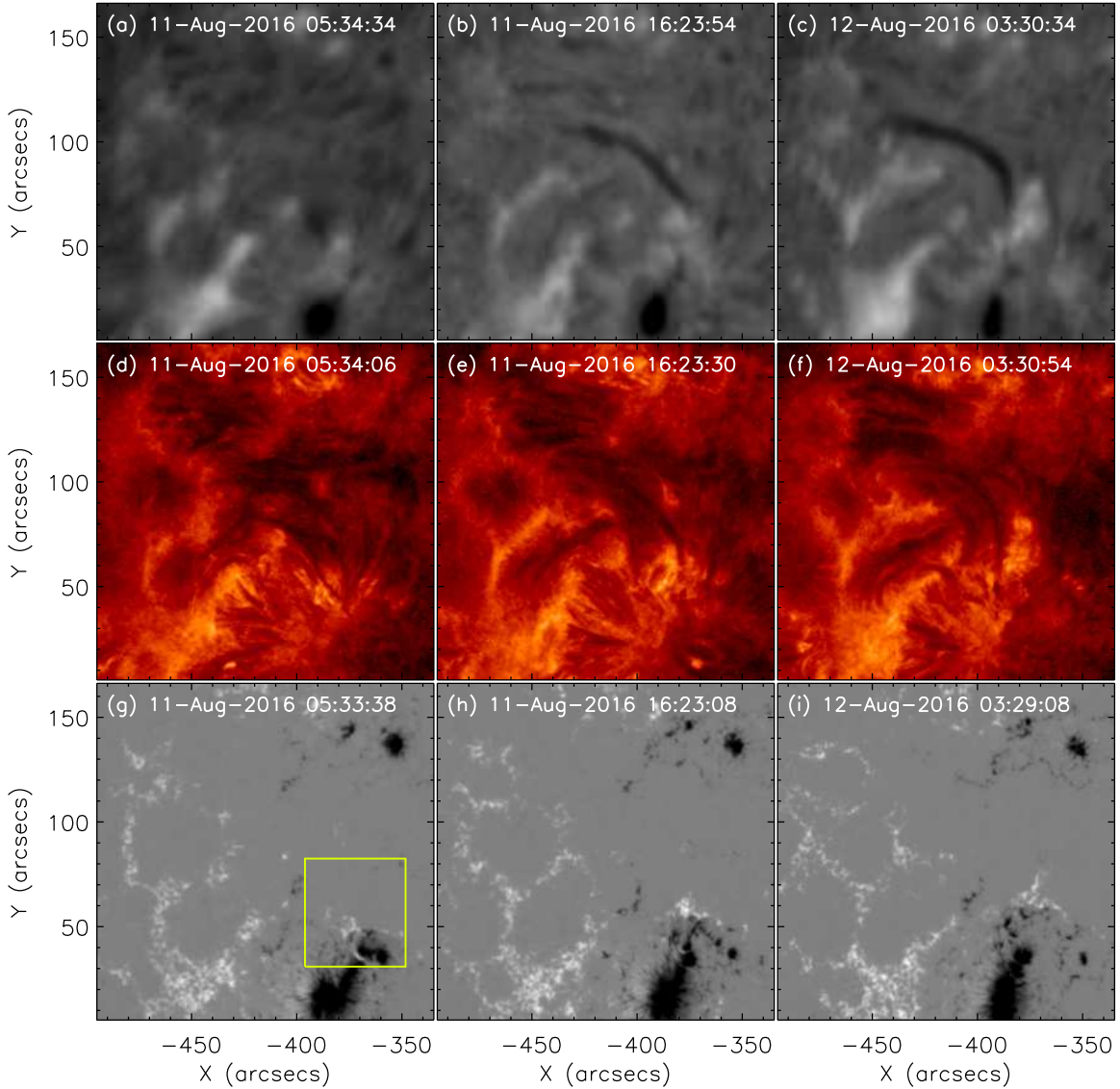}
\caption{Formation process of the filament. (a)-(c) H$\alpha$ images observed by GONG in different moments. (d)-(f) The corresponding SDO/AIA 304 $\rm \AA$ images. (g)-(i) The corresponding line-of-sight magnetic fields from SDO/HMI. White patches denote the magnetic field with positive polarity, while black ones denote the magnetic field with negative polarity. The yellow rectangle in panel (g) indicates the region for calculating different physical quantities in Fig.\ref{figure3}. \label{figure1}}
\end{figure}

A filament, located in active region NOAA 12574 on northeast hemisphere (see Fig.\ref{figure0} (a)), formed gradually during the period from 04:00 UT on August 11 to 04:00 UT on August 12, 2016. The entire formation process of the filament was observed by several ground-based and spaced-based telescopes. Fig.\ref{figure1} displays the formation process of the filament in SDO/AIA 304 $\rm\AA$ and H$\alpha$ bands. As is shown by GONG H$\alpha$ observations in panels (a)-(c), the filament was absent at 05:34:34 UT on August 11. The filament formed completely at 03:30:34 UT on August 12. The corresponding SDO/AIA 304 $\rm \AA$ images are shown in panels (d)-(f). Panels (g)-(i) show the photospheric line-of-sight magnetic fields observed by SDO/HMI. The detailed formation process of this filament is shown in the animated version of Fig.\ref{figure1} (a).

\begin{figure}[ht!]
\figurenum{3}
\plotone{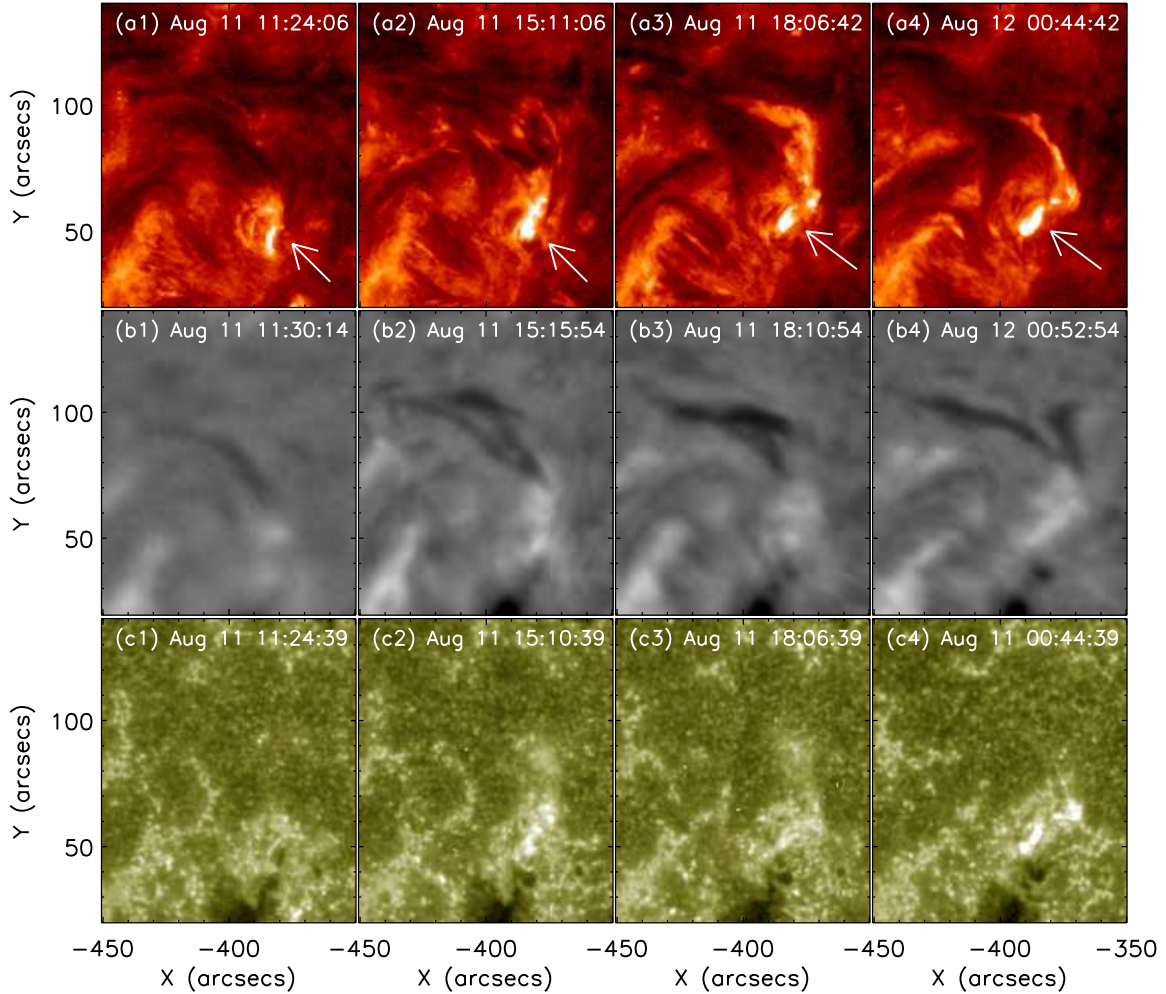}
\caption{A series of typical jets occurred on the western foot-point of the filament, which injected massive plasma into the filament. (a1)-(a4) The jets in SDO/AIA 304 $\rm\AA$ images. The white arrows indicate jets related to the formation of the filament. (b1)-(b4) GONG H$\alpha$ observations after each jet. (c1)-(c4) The corresponding SDO/AIA 1600 $\rm\AA$ images during each jet.\label{figure2}}
\end{figure}

According to the animation of Fig.\ref{figure1} and Fig.\ref{figure2}, a series of jets occurring on the western foot-point of the filament during the formation of the filament, injected mass into the coronal height. As cool plasma was injected into the filament by the jets, the dark and elongate filament appeared in the field of view. During the early period from 04:00 UT to 12:30 UT on August 11, most of plasma was lifted to the filament height from the western foot-point by the jets, and then threw down to the other foot-point. There was only a fraction of plasma that could be retained in the filament height. This might be related to the local magnetic structure, which is too little ``dips" or twisted magnetic structure existed in the local corona to capture the injected plasma. This also means that not all plasma injected by the jets can be transformed to the filament material. At a later period, most of the lifted plasma was trapped in the filament and become the material of the filament instead of falling down to the other foot-point. Fig.\ref{figure2} shows serval jets occurring in the vicinity of the western foot-point of the filament. Panels (a1)-(a4) show four jets in different moments in SDO/AIA 304 $\rm \AA$ observations, which are marked by white arrows nearby the western foot-point of the filament. Panels (b1)-(b4) exhibit that the cool material was injected into the filament from low solar atmosphere after each jet, while the brightening in SDO/AIA 1600 $\rm\AA$ could also be identified nearby the western foot-point of the filament during each jet in panels (c1)-(c4). It is found that a number of cool plasma blobs were directly injected into the filament driven by each jet.

\begin{figure}[ht!]
\figurenum{4}
\plotone{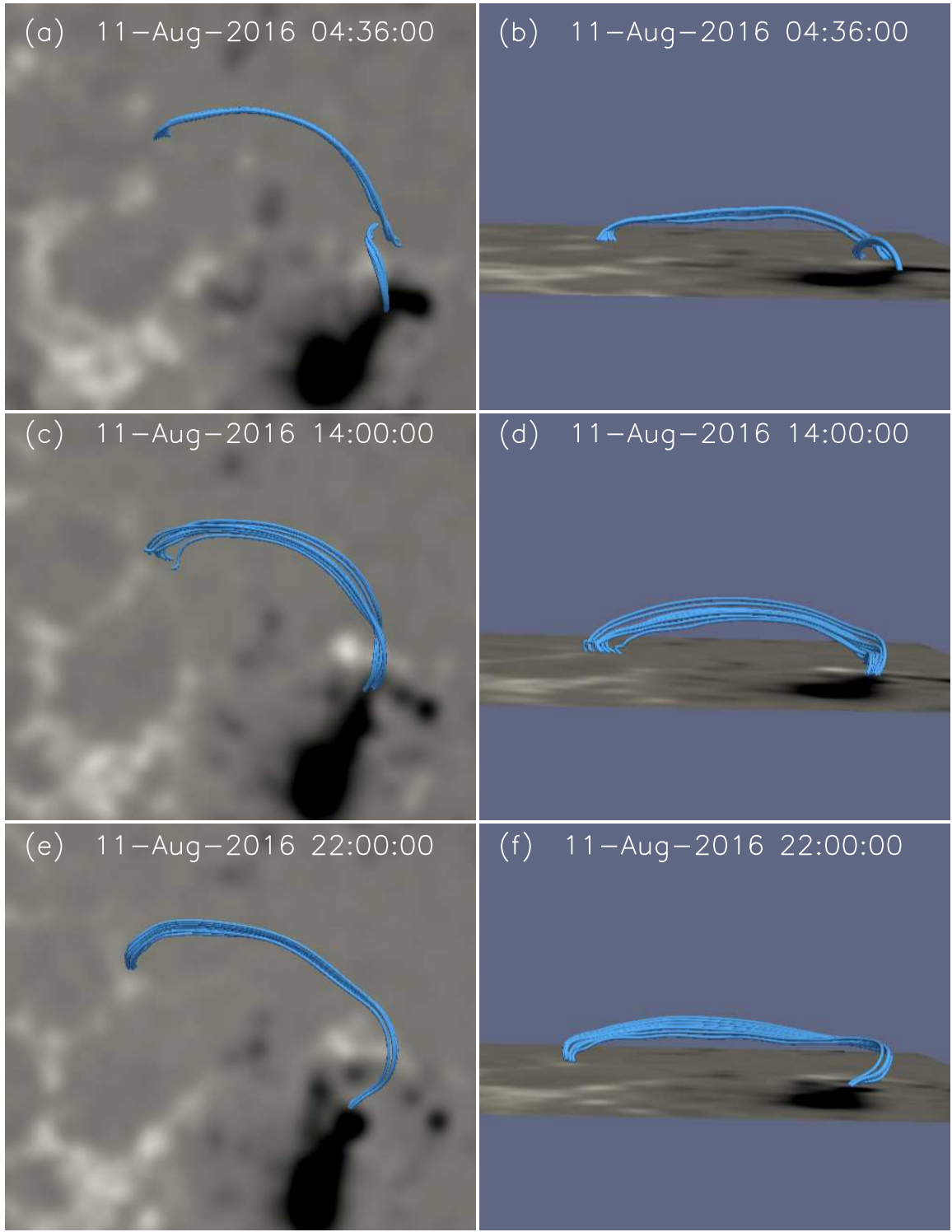}
\caption{The selected  magnetic field lines derived by using NLFFF extrapolation model at different times and different views. The left columns are seen from top side view, while the right columns are seen from left-side view. The blue lines indicate the selected magnetic field lines. The background indicates the radial magnetic field. \label{figure9}}
\end{figure}

As described in section \ref{sec:obser methods}, we use the NLFFF model to extrapolate the magnetic structures of the jets and the filament. Fig.\ref{figure9} shows the selected magnetic field lines at different moments. Panels in the left column show the magnetic structure seen from top side view, while panels in the right column show the perspective from the left (solar East). Panels (a) \& (b) and panels (c) \& (d) show the magnetic field lines at 04:00:00 UT and 14:00:00 UT on August 11, respectively, while panels (e) \& (f) show the selected magnetic field lines at 22:00:00 UT on August 11. At 04:36:00 UT, there were two flux tubes existing in the field of view before eruption of jets. The small tube corresponds to the magnetic structure of the emerging magnetic field. When the jet erupted, it reconnected with the big flux tube and formed much larger flux tube (see Fig.\ref{figure9} (c) and \ref{figure9} (d)). The repeated process was observed during a series of jet eruptions, and the magnetic structure of the filament formed finally (see Fig.\ref{figure9} (e) \& (f)).

\begin{figure}[ht!]
\figurenum{5}
\plotone{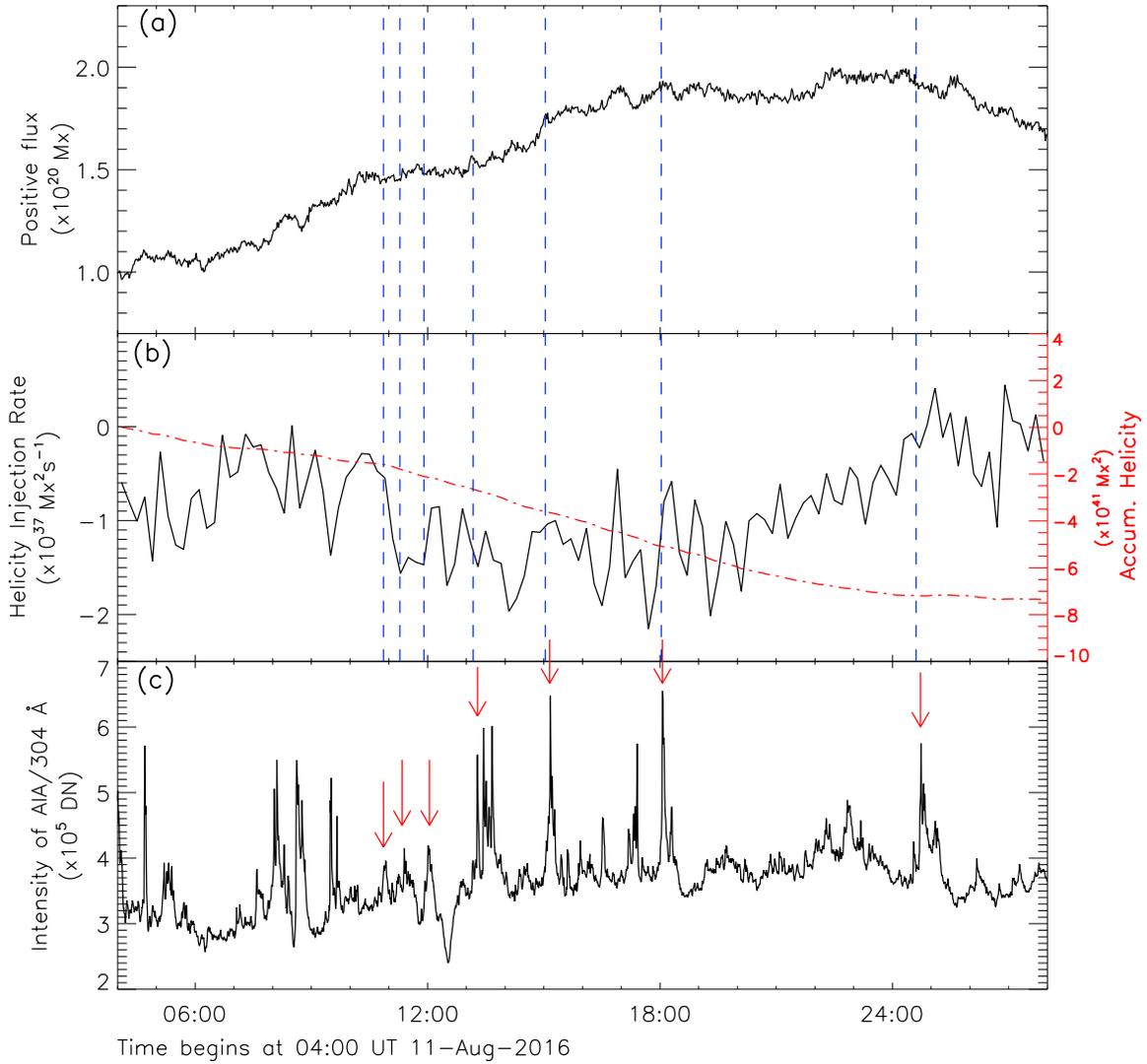}
\caption{Time variations of the positive magnetic flux, magnetic helicity and SDO/AIA 304 $\rm\AA$ intensity in the yellow rectangle of Fig.\ref{figure1} (g) during the period from 04:00 UT August 11 to 04:00 UT August 12, 2016. (a) The time variation of the positive magnetic flux. (b) The time variations of the helicity injection rate and accumulated helicity. Helicity injection rate is shown by the solid black line and the red dotted-dashed line indicates the accumulated helicity. (c) The line profile of the time variation of the intensity of SDO/AIA 304 $\rm\AA$. The perpendicular red arrows denote several jets related to the material injection of the filament nearby the western foot-point of the filament. Blue dash lines indicate the onset of each jet.\label{figure3}}
\end{figure}

In order to understand the physical mechanism of these jets injecting material for the filament, we calculate the magnetic flux, magnetic helicity and intensity of SDO/AIA 304 $\rm \AA$ nearby the western foot-point of the filament (the region marked by the yellow box of the Fig.\ref{figure1} (g)). We only calculate the positive magnetic flux because of the complexity of the negative magnetic flux in this region. Panels (a), (b) and (c) of Fig.\ref{figure3} show the time variations of the positive magnetic flux, magnetic helicity and intensity of SDO/AIA 304 $\rm \AA$ during the period from 04:00 UT on August 11 to 04:00 UT on August 12 in the yellow box of the Fig.\ref{figure1} (g), respectively. The increases of the SDO/AIA 304 $\rm\AA$ intensity during the periods of the jets related to the formation of the filament nearby the western foot-point were identified, which are marked by red arrows in the panel (c). Vertical blue dashed lines indicate the onsets of the jets. In the panel (a), it is found that the positive magnetic flux almost increased to twice during the period from 04:00 UT on August 11 to 00:00 UT on August 12. The increase of the magnetic flux manifests that the magnetic flux emerged in the vicinity of the western foot-point of the filament. It is believed that the flux emergence nearby the western foot-point of the filament played an important role in the occurrences of these jets. On the other hand, the magnetic flux decreased by 5-10$\%$ after each several jets (such as the last four marked jets). Panel (b) shows the evolution of helicity injection rate and accumulated helicity. The helicity injection rate is shown by the solid black line, while the red dotted-dashed line indicates the accumulated helicity. The accumulated helicity is calculated by the time integral of the helicity injection rate and we set it to be zero at 04:00 UT on 2016 August 11. It is found that the negative helicity injection was dominant and the negative accumulated helicity was constantly increasing. It is noted that the negative helicity is injected into the upper atmosphere nearby the western foot-point of the filament. It does not mean that all the injected helicity is stored in the filament. Based on above observations, we propose that these jets were triggered by the magnetic reconnection between closed pre-existing magnetic fields and emerging magnetic fields nearby the western foot-point of the filament. Moreover, these jets injected massive cool plasma into the filament. Some of the lifted plasma could stay in the corona height and became the filament material. However, some of the lifted plasma could not maintain stability in the corona and eventually descended at the other foot-point. Thereafter, the dark and elongated filament appeared. In the meantime, some of the post-reconnected magnetic fields would sink because of the relaxation of magnetic fields after these jets, which corresponded to the \textbf{cancellation} of the magnetic flux in the photosphere.

\subsection{Two material injection events related to the filament formation} \label{jets}

To better investigate the material injection of this active region filament, we present two material injection events (Jet A and Jet B) which are studied in detail as examples, respectively. Jet A occurred at 18:02 UT on August 11, and Jet B occurred at 00:42 UT on August 12. We utilize the data observed by SDO/AIA, GST/BBSO and Hinode/SP to investigate the Jet A, while the data observed by SDO/AIA, DST/Hida and IRIS were used to investigate the Jet B.

\subsubsection{The Jet A at 18:02 UT on August 11 \label{jet1802}}

\begin{figure}[ht!]
\figurenum{6}
\plotone{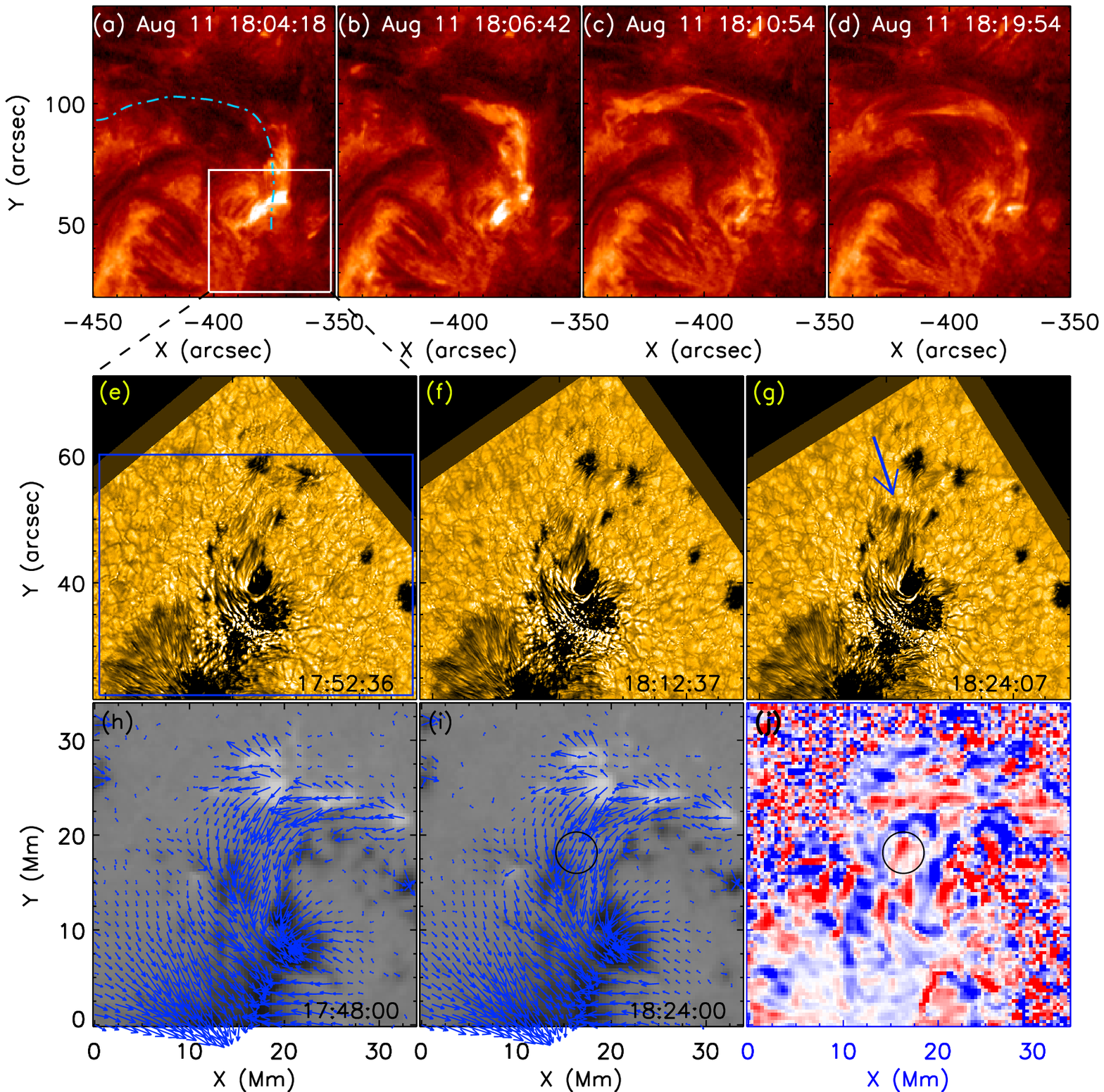}
\caption{Evolution of a jet at 18:02 UT on August 11, 2016. (a)-(d): SDO/AIA 304 $\rm \AA$ images. The pink dotted-dashed line in the panel (a) denotes the path of the slice, which the white box denotes the field of view of panels (e)-(j). (e)-(g): Tio images observed by GST/BBSO. The blue box in the panel (e) denotes the field of view of Figs.\ref{figure5} (b) and (c), while the blue arrow in the panel (g) point out the increase of the dark thread after the jet. (h)-(i): Vector magnetograms observed by SDO/HMI. The blue arrows denote the transverse magnetic fields, while the backgrounds denote the vertical magnetic fields. (j) The different magnetic field inclination between 18:24:00 UT and 17:48:00 UT. The red color denotes positive value, while the blue color denotes negative value. Two black circles mark the location of the increase of the dark thread after the jet. \label{figure4}}
\end{figure}

The Jet A occurred at 18:02 UT on 2016 August 11. Panels (a)-(d) of Fig.\ref{figure4} show the process of the jet in SDO/AIA 304 $\rm\AA$ images. As is shown, the jet occurred nearby the western foot-point of the filament, and then lifted the plasma from the western foot-point of the filament into the filament. In order to investigate the photospheric response of this jet, TiO images observed by GST are utilized to show the change of photosphere during the jet. The field of view of TiO images corresponds to the region marked by the white box in the panel (a). Panels (e)-(g) of Fig.\ref{figure4} exhibit TiO observations at the western foot-point of the filament during the period covering the jet. It is found that some dark threads in the photosphere appeared in the vicinity of the western foot-point after the jet, which was marked by the blue arrow in the panel (g). Panels (h)-(i) of Fig.\ref{figure4} show the vector magnetograms observed by SDO/HMI at 17:48:00 UT and at 18:24:00 UT, which correspond to the times before and after the jet, respectively. Panel (j) of Fig.\ref{figure4} shows the difference of magnetic field inclination at 18:24:00 UT and at 17:48:00 UT. The magnetic field inclination is calculated by the follow formula: $\theta = arctan(\sqrt{bx^2+by^2}/|bz|)$. Thus, the positive value of the difference of magnetic field inclination means that the magnetic field became more horizontal, while the negative value means that the magnetic field became more vertical. It is found that the difference value of magnetic inclination at the location marked by the black circle in the panel (j) of Fig.\ref{figure4} is positive where some dark threads appeared after the jet. This means that the magnetic field became more horizontal after the jet. Therefore, it is reasonable to suspect that the appearance of dark threads would be associated with the change of the magnetic field after the jet. Due to the magnetic reconnection between the pre-existing magnetic field and emerging magnetic field during the jet, the magnetic field became more horizontal with the relaxation or sinking of the post magnetic reconnection field. Therefore, the more dark threads appeared as the magnetic field became more horizontal. In other words, dark threads are the representation of the magnetic field with big inclination in the photosphere.

\begin{figure}[ht!]
\figurenum{7}
\plotone{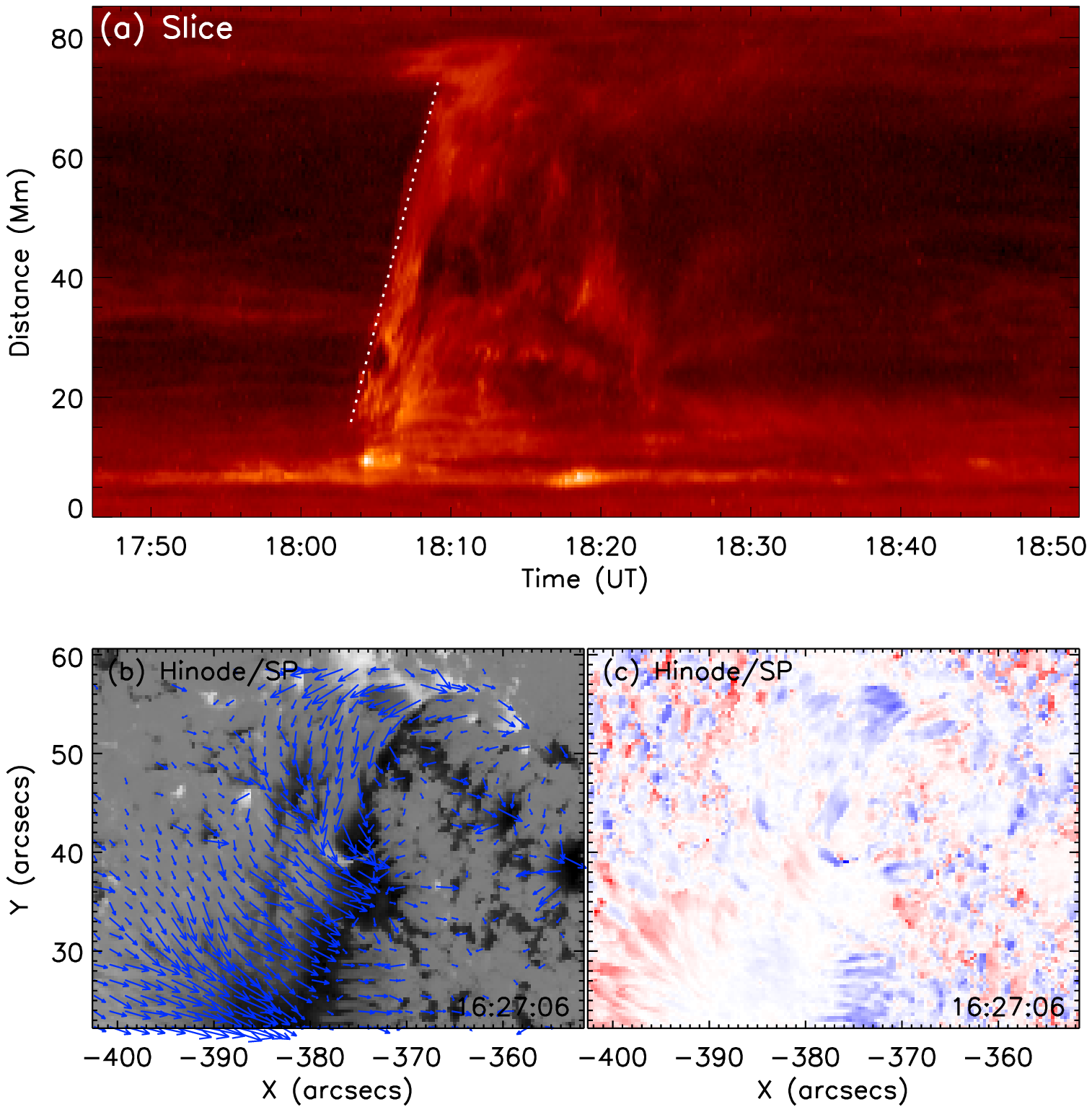}
\caption{(a) The time-distance diagram along the pink dot-dash line in the panel (a) of Fig.\ref{figure4}. (b) The vector magnetic field observed by the Hinode/SP instrument at the western foot-point of the filament. The background denotes the radial magnetic field while blue arrows denote transverse magnetic field. (c) The doppler shift derived from FI 6302.28 $\rm \AA$ observed by Hinode/SP instrument.\label{figure5}}
\end{figure}

In order to understand the property of the injected plasma, we make a time-distance diagram to estimate the velocity of the injected plasma along the axis of the filament. The path of the time-distance diagram is made by the pink dotted-dashed line in the panel (a) of Fig.\ref{figure4}. Fig.\ref{figure5} (a) shows the time-distance diagram derived from the SDO/AIA 304 $\rm \AA$ observations. According to the time-distance diagram, the projection velocity of the heated plasma along the axis of the filament is about 162.6$\pm$5.4 $km/s$. Figs.\ref{figure5} (b) and \ref{figure5}(c) show the vector magnetic field and Doppler shift on the photosphere nearby the western foot-point of the filament. The field of view of panel (b) of Fig.\ref{figure5} is marked by the blue box in the panel (e) of the Fig.\ref{figure4}. The blue arrows in the panel (b) denote the transverse magnetic fields, while the white and the black patches denote the radial positive and negative magnetic field, respectively. It is found that the western foot-point of the filament rooted in negative magnetic polarity. Therefore, it is easy to derive that the direction of the injected plasma along the axis of the filament was from negative polarity to positive one.

\subsubsection{The Jet B at 00:42 UT on August 12} \label{0042}
\begin{figure}[ht!]
\figurenum{8}
\plotone{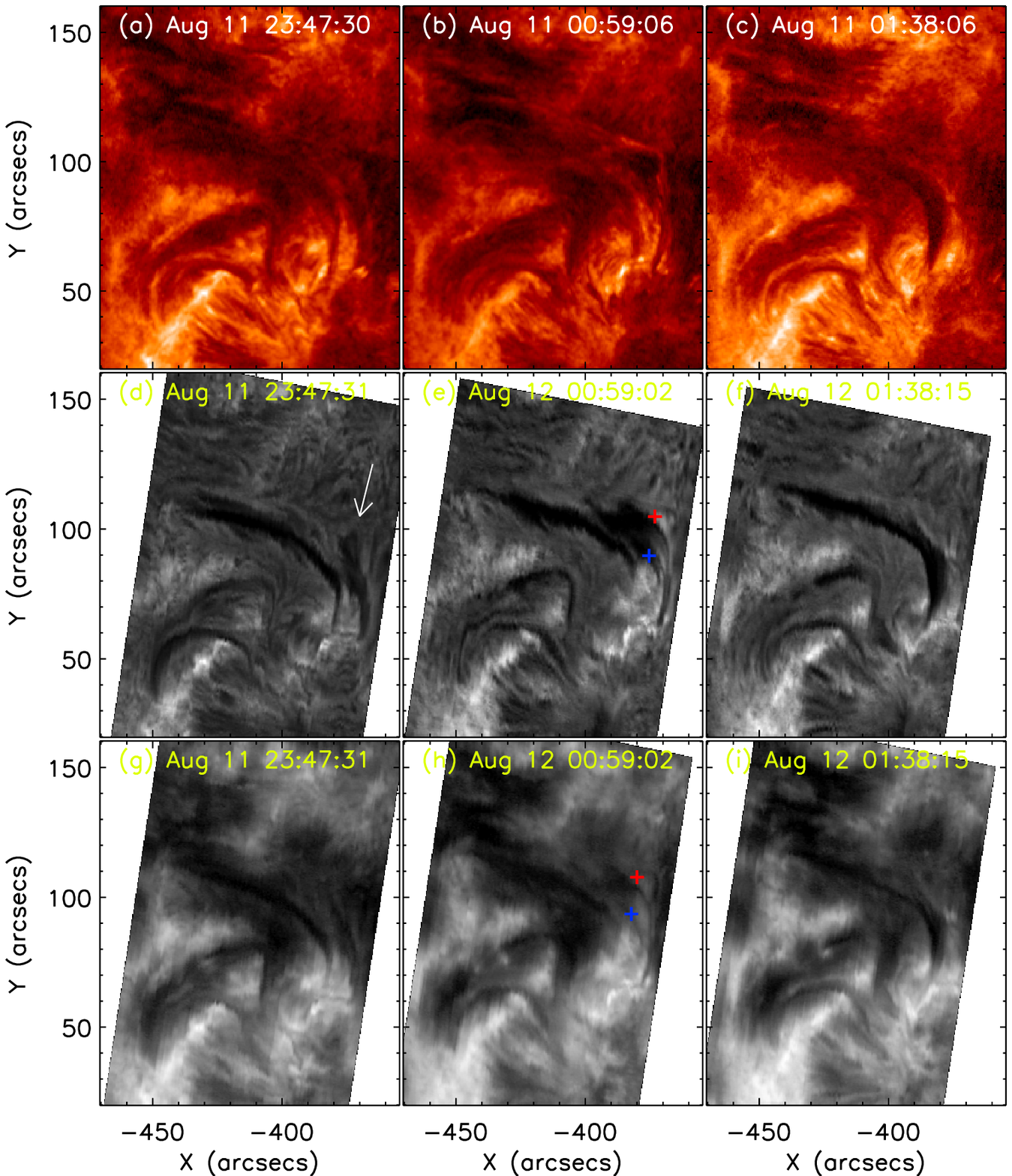}
\caption{Observations of a jet at 00:42 UT on August 12. (a)-(c) SDO/AIA 304 $\rm \AA$ images before, during and after the jet. (d)-(f) Corresponding constructed H$\alpha$ images by the center of H$\alpha$ spectrum observed by Hida/DST. The white arrow in the panel (d) indicates the dark structure nearby the western foot-point of the filament. The blue and red plus sign (+) in the panel (e) are the sites of the spectral lines in Fig.\ref{figure7} (a), respectively. (g)-(i) Corresponding constructed Ca II K images by the center of Ca II K spectrum observed by Hida/DST. The blue and red plus sign (+) in the panel (h) are sites of the spectral lines in Fig.\ref{figure7} (b), respectively.  \label{figure6}}
\end{figure}

The Jet B occurred at 00:42 UT on August 12. This jet also appeared in the vicinity of the western foot-point of the filament at the same place as other jets. This material injection process was captured by DST/Hida, SDO/AIA and IRIS telescopes, simultaneously. Fig.\ref{figure6} shows the process of the jet in different wavelengths and in different moments, respectively. Panels (a)-(c) show the jet acquired at SDO/AIA 304 $\rm \AA$, while Panels (d)-(f) and (g)-(i) are reconstructed images by using the center of the H$\alpha$ 6562.8 $\rm \AA$ and Ca II K 3933 $\rm \AA$ spectrum observed by the DST/Hida, respectively. According to the Fig.\ref{figure6}, massive cool plasma was injected into the filament driven by this jet in different wavelengths. After the jet, the filament became broader and darker. At about 23:47 UT, the dark structure (marked by the white arrow in the panel (d)) filled with massive plasma, existed nearby the western foot-point of the filament. At about 00:59 UT, the plasma in the dark structure started to be injected into the filament driven by the jet at 00:42 UT. At about 01:38 UT, all of the plasma had injected into the filament. And then, this dark structure (marked by the white arrow in the panel (d)) disappeared while the filament became broader and darker.

\begin{figure}[ht!]
\figurenum{9}
\plotone{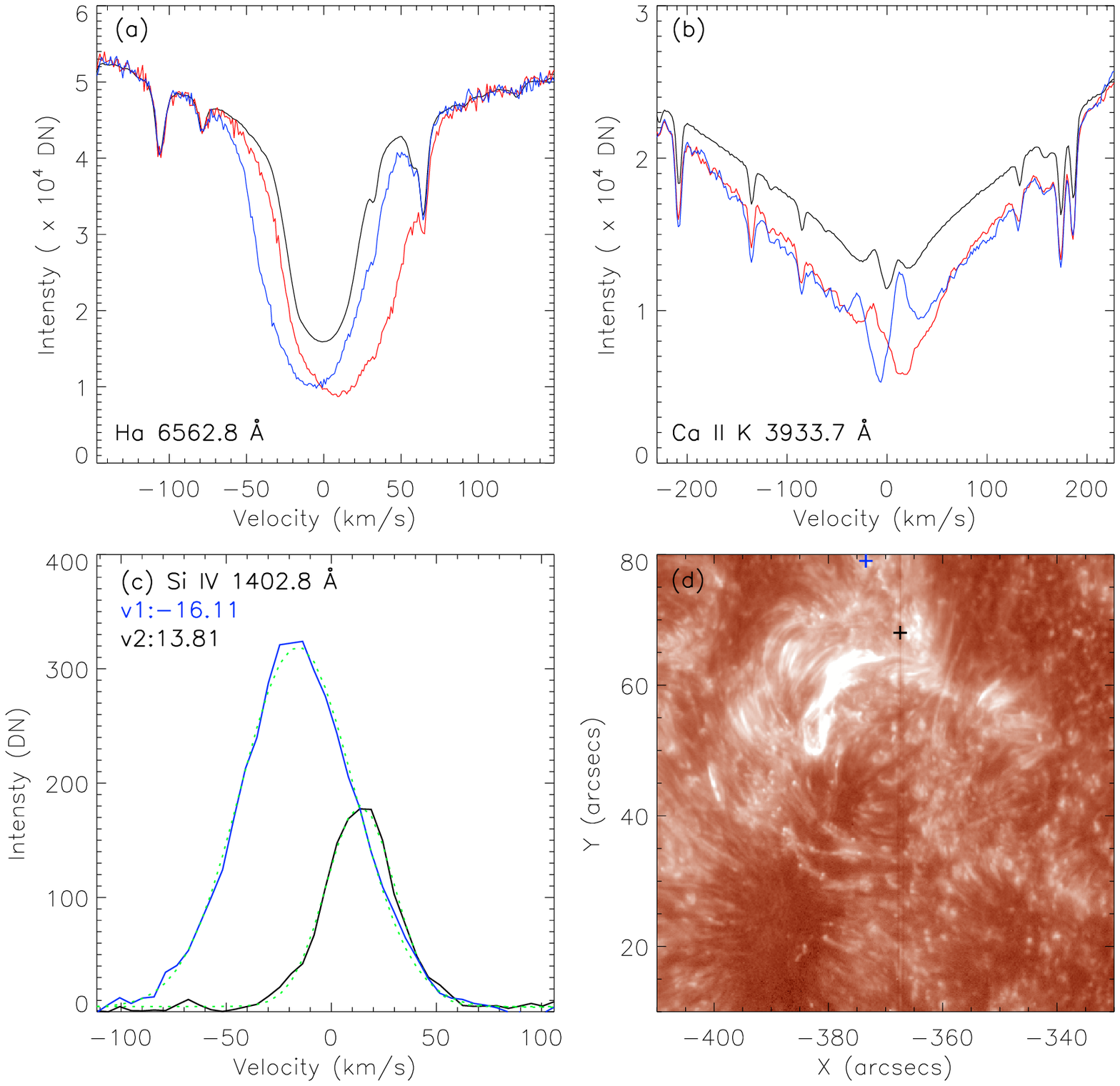}
\caption{(a) Line profiles of H$\alpha$ from DST/Hida. The back line denotes the line profile of quiet Sun. The blue one sites in the blue plus sign (+) in the panel (e) of Fig.\ref{figure6}, while the red one sites in the red plus sign (+). (b) Line profiles of Ca II K from DST/Hida. The back line denotes the line profile of quiet Sun. The blue one sites in the blue plus sign in the panel (h) of Fig.\ref{figure6}, while the red one sites in the red plus sign. (c) Line profiles of Si IV 1402.8 $\rm\AA$. The blue line sites at blue plus sign in the panel (d), which the black line is corresponding to the black plus sign in the panel (d). (d) The SJI of 1400 $\rm\AA$ observed from IRIS.\label{figure7}}
\end{figure}

In order to investigate the injected plasma, we analyze the spectrum of H$\alpha$ and Ca II K line during the period of the jet. Fig.\ref{figure7} (a) exhibits the profiles of H$\alpha$ at different sites. The blue line is the profile of H$\alpha$ at the site marked by the blue plus sign (+) in the panel (e) of Fig.\ref{figure6}, while the red line is the profile of H$\alpha$ at the site marked by the red plus sign in the panel (e). The black line indicates the profile of the quiet Sun. The blue line displays blue shift signature while the red one displays red shift signature. Fig.\ref{figure7} (b) shows the line profiles of the Ca II K line at different sites. The blue line shows the profile of Ca II K at the site marked by the blue plus sign in the panel (h) of Fig.\ref{figure6}, while the red line is the one at the site marked by the red plus sign in the panel (h). The black line indicates the profile of the quiet Sun, which is the same as the panel (a). The similar feature as H$\alpha$ profile could be found in Ca II K line, in which the blue and red shifts are also found in the blue and red lines, respectively. These blue and the red shifts manifest that the cool injected plasma displayed different motion states at different site during the jet.

Panel (c) of Fig.\ref{figure7} shows the profiles of Si IV 1402.8 $\rm \AA$ observed by IRIS. The blue line shows the profile of Si IV line at the site marked by the blue plus sign in the panel (d) at 01:02:26 UT on August 12, while the black line shows the profile of Si IV line at the site marked by the black plus sign at 01:06:49 UT on August 12. The panel (d) shows the SJI of the 1400 $\rm \AA$ band nearby the western foot-point of the filament during the jet. With the method described in the section 2, we use a single Gaussian fitting method to derive the Doppler velocities of these two profiles of Si IV lines. The Doppler velocity of the blue line was about -16.11 km/s, while the one of the black line was about 13.81 km/s. This means that the heated plasma driven by the jet at the position of the blue plus sign showed a upward motion, but the one at the position of the black plus sign exhibited downward motion. As is well known, some plasma were injected into the filament, which is manifested as blueshifts of the Si IV line. However, some injected plasma could not remain in the filament owing to the imbalance between magnetic support and gravitation, and fell down to the western foot-point of the filament. Therefore, it showed the red shift in the line profile of Si IV 1402.8 $\rm \AA$ at 01:06:49 UT on August 12.

\begin{figure}[ht!]
\figurenum{10}
\plotone{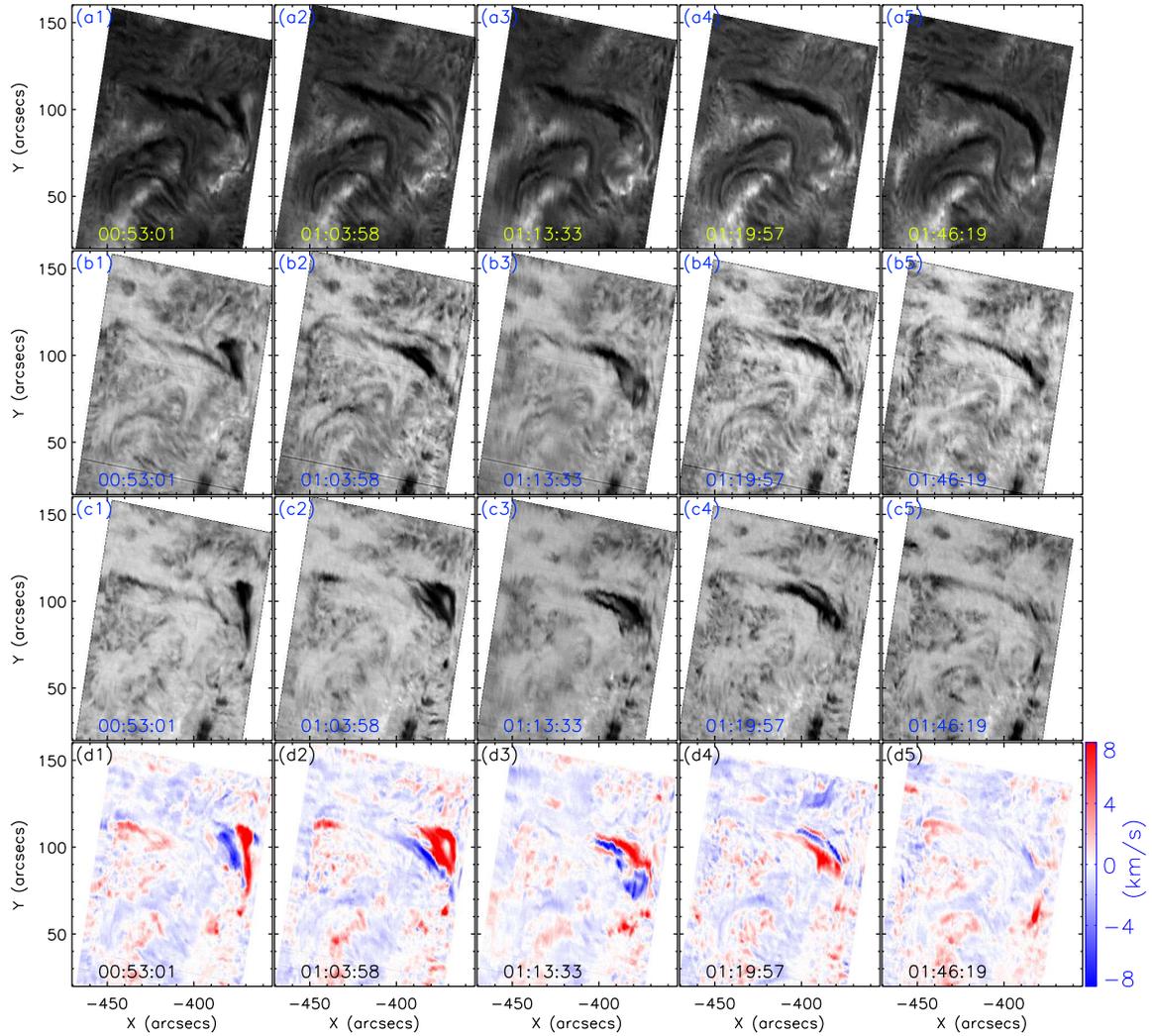}
\caption{Evolution of the jet at 00:42 UT on August 12 observed from DST/Hida. (a1)-(a5) The constructed images of center of H$\alpha$. (b1)-(b5) The constructed images of H$\alpha$+0.6 $\rm\AA$. (c1)-(c5) The constructed images of H$\alpha$-0.6$\rm\AA$. (d1)-(d5) The corresponding Doppler shifts derived by H$\alpha$ spectra. \label{figure8}}
\end{figure}

Fig.\ref{figure8} shows the process of the jet in H$\alpha$ line observed by DST/Hida. Reconstructed images at five different moments during the jet are shown. The reconstructed image is constructed from a set of one scan data at the fixed wavelengths. Panels (a1)-(a5) show the jet in the center of the H$\alpha$, while panels (b1)-(b5) and panels (c1)-(c5) are the corresponding images reconstructed by using +0.6 $\rm\AA$ and -0.6 $\rm\AA$ of H$\alpha$ spectrum from DST/Hida, respectively. Panels (d1)-(d5) show the Doppler velocity derived from H$\alpha$ spectral data with the method described at Section 2. It is noted that the maximum of the Doppler velocity is smaller than 15 km/s. In panel (a1), massive dark plasma was driven by the jet to move close to the filament at 00:53:01 UT. In the meanwhile, panels (b1) and (c1) exhibit the corresponding images constructed in H$\alpha$-0.6 $\rm\AA$ and H$\alpha$+0.6 $\rm\AA$. The doppler velocity is shown in the panel (d1). As is shown, the eastern part of the injected plasma exhibited blue shift, while the western part of the injected plasma showed red shift. Combined with the observation of SDO/AIA 304 $\rm\AA$, we infer that the injected plasma was rotating. Based on the location of the filament, it was anticlockwise when viewed from the apex to the western foot-point of the filament.

At 01:03:58 UT on August 12, the jetted plasma began to be injected into the filament. At 01:13:33 UT, the jetted plasma had been injected to the filament and became the material of the filament. At 01:19:57 UT, the filament experienced a little active after the injection of material. At 01:46:19 UT, the filament became more broader and larger after this event. According to the corresponding velocity in the panels (d2)-(d5), the two interesting features were found. Firstly, as the rotated plasma were injected to the filament, the filament also experienced anticlockwise rotation which showed in the panels (d2) and (d3). Secondly, the filament appeared inverse rotation after the anticlockwise rotation. We conjecture that the anticlockwise rotation of the filament was caused by the injected twisted magnetic field structure which was transformed from the jet into the filament. And the reason of the inverse rotation of the filament was due to the back action which balanced the magnetic field configuration after the injection of the twisted magnetic field by the jet. According to these behaviors above, we conclude that the jet caused by the reconnection between the twisted structure magnetic field with massive plasma and the filament magnetic field also injected the magnetic helicity into the filament. In other words, the jet not only forced the plasma into the filament, but also effected the configuration of the filament magnetic field.
\section{Estimation of the jets mass, the filament mass and the upper limit of the twist}
With the assumptions that the speed of the injected plasma is constant and the cross-section of the jet is circular, we calculate the total mass carried by the jets. The quality of mass injection can be calculated by the following equation:
\begin{equation}\label{jetsmass}
  M_j =m_Hn_H\frac{w^2\pi}{4}vt
\end{equation}
where $m_H$ is the quality of the hydrogen atom, $n_H$ is the total hydrogen number density, $w$ is the jet width, $v$ is the jet speed, $t$ is the jet duration. We assume that the density of the jets equal to that of the chromosphere. The total hydrogen number density $n_H$ is about $(1.71+1.55)\times 10^{10}$ $cm^{-3}$ (the model A from \cite{fon93}), while the mass of the hydrogen atom $m_H$ is about $1.67\times 10^{-24}$ $g$.
\begin{deluxetable*}{cccccc}
\tablecaption{The dynamical parameters of jets}
\tablecolumns{6}
\tablenum{1}
\tablewidth{0pt}
\tablehead{
\colhead{Start Time} &
\colhead{Jet Width($w$)} &
\colhead{Jet Speed($v$)} &
\colhead{Duration($t$)} &
\colhead{Mass($M_j$)} \\
\colhead{(UT)} &
\colhead{(Mm)} &
\colhead{(km/s)} &
\colhead{(minutes)} &
\colhead{($10^{14}g)$}
}
\startdata
2016-Aug-11 10:52 & 0.97 & 73.2 & 11 & 0.02 \\
2016-Aug-11 11:28 & 2.04 & 55.3 & 14 & 0.08 \\
2016-Aug-11 11:57 & 2.70 & 133.2 & 16 & 0.40 \\
2016-Aug-11 13:17 & 6.88 & 174.6 & 33 & 6.99 \\
2016-Aug-11 15:02 & 7.57 & 121.9 & 21 & 3.76 \\
2016-Aug-11 18:02 & 4.94 & 162.6 & 22 & 2.24 \\
2016-Aug-12 00:42 & 5.59 & 88.7 & 35 & 2.49 \\
\enddata
\end{deluxetable*}
Table 1 presents the parameters of the jets. The total mass carried by jets is about $16\times 10^{14}$ $g$. The last four jets supply the most of the mass, which are more than 97$\%$ of the total mass. According to the enhancements of the SDO/AIA 304 $\rm\AA$ intensity during the jets (see Fig.\ref{figure3} (c)), the intensity levels of the last four jets are bigger than the former three jets. Therefore, it is reasonable that the last four jets carried the most of the mass.

Under a simple assumption that the filament is circular column, the quality of the mass in the filament $M_f$ is estimated using the equation: $M_f=n_Hm_H\frac{w^2_f\pi}{4}L_f$, where $n_H$ is the total hydrogen density of the filament, $w_f$ is the filament width, and $L_f$ is the filament length. According to the shape of the filament, we obtain $L_f=56.1$ Mm and $w_f=10.4$ Mm. The total hydrogen number density $n_H$ in the filament is between $3\times 10^{10}$ $cm^{-3}$ \citep{ste97} and $3\times 10^{11}$ $cm^{-3}$ \citep{hir86}. The total filament mass is estimated to be in the range $(2.39-23.9)\times10^{14}$ $g$. Comparing the total mass carried by the jets with the mass of the filament, the total mass carried by the jets is compatible with the range of the filament mass and very closes to the upper value of the filament mass. Thus, the estimated mass loading by the jets is sufficient to account for the mass in the filament.

We estimate the twist injected into the filament system by the formula, $T=H/\phi^2$, where $H$ is the cumulative helicity and $\phi$ is the magnetic flux contained in the eventual filament. $T$ presents the twist in the units of turns. It would provide an upper limit to the number of windings in the filament, if all the helicity injected were stored in the filament. The flux in the filament can be estimate using by the equation: $\phi=\int_s B_tds$, where $B_t$ is the transverse magnetic field strength, $s$ is the cross-section of the filament. The range of the transverse magnetic fields strength in the active-region filament is from 500 to 600 G \citep{kuc09,xu12}. The cross-section of the filament, $s=(w_f/2)^2*\pi$, is about $8.5\times10^{17}$ $cm^2$. Using these values, we derive the magnetic flux in the eventual filament is from $4.25 \times 10^{20}$ to $5.10\times10^{20}$ $Mx$. According to the formulas, the upper limit of the twist $T$ is between 2.28 and 4.11 turns.
\section{Conclusion and discussion}\label{sec:conclusion}
In this study, we examined the formation of a filament in NOAA AR 12574. The material injection of the filament is mainly investigated. This is a rare set of observations. Beginning with the time when the filament did not exist, both ground-based and space borne observatories provided continuous coverage of its formation over the next two days. SDO, GONG and Hida observations covered the entire process of the filament material injection. We focus on the mechanism of the transportation of the filament material. On the one hand, we analyze the variation of the magnetic flux nearby the western foot-point of the filament, which is associated to the material injection events (jets). On the other hand, two mainly material injection events are investigated in detail. One occurred at 18:02 UT on August 11 and the other occurred at 00:42 UT on the next day. The main results are as follows:

1. Material of the filament originate from the low solar atmosphere. The material of the filament are supplied by a series of jets. These jets occurred nearby the western foot-point of the filament and forced massive plasma from low atmosphere to the filament. Jets provide a sufficient and direct way to carry dense and cool plasma from the low atmosphere into high atmosphere.

2. Flux emergence should be responsible for occurrences of these jets. These jets were caused by the magnetic reconnection between the emerging magnetic field and pre-existing magnetic field nearby the western foot-point of the filament.

3. The projection velocity of heated plasma along the filament axis during the Jet A at 18:02 UT on August 11 was about 162.6$\pm$5.4 km/s. For the Jet B on 00:42 UT on August 12, we find the LOS velocity of cool injected plasma derived by H$\alpha$ spectrum was smaller than 15 km/s. Furthermore, Using the Si IV line from IRIS observations, it is found that the blue shift was about 16.11 km/s while the red shift was about 13.81 km/s.

4. Jets not only injected the plasma from the low atmosphere into the filament, but also injected the magnetic helicity into the filament, simultaneously. The rotating injected plasma manifested that the twisted structure had existed in the emerging magnetic field. The twist in the emerging magnetic field could be transformed into the filament when the emerging magnetic field interacted with the filament.

5. The total mass carried by jets is estimated to be about $16\times 10^{14}$ $g$, while the filament mass is estimated to be in the range $(2.39-23.9)\times10^{14}$ $g$. Thus, the mass carried by the jets is sufficient to account for the mass in the filament.

Previous work on quiescent filaments have shown that the mass in such filaments is supplied from the low atmosphere instead of the corona \citep{pik71,zir94,mac10}. The same origin can be also inferred for the active region filament in our study. This study suggests that jets can provide the supply of mass from the lower atmosphere to active region filaments, consistent with previous reports \citep{wan99,cha03,liu05,zou16}. Although the evaporation-condensation model provides an another way to supply the mass for the filaments, which is supported by numerous numerical simulations, but the observation evidences are rare \citep{mac10,liu12,par14}. Besides, there are two different magnetic reconnection sites for injecting material into the filament in the previous studies. One is at the PIL, and the other is nearby the foot-points of the filament \citep{cha03,liu05}. In this study, the magnetic reconnections occurred at the western foot-point of the filament.

In this study, a lot of cool material was carried into the filament by the jets. \cite{tak13} studied the acceleration mechanisms of chromospheric jets associated with emerging flux using a two-dimensional magnetohydrodynamic (MHD) simulation and found that slow-mode waves play key roles in the acceleration mechanisms of chromospheric jets. They mainly investigated two example jets resulting from magnetic reconnection near the photosphere and  in the upper chromosphere, and suggested three types of acceleration mechanisms of cool jets: Shock acceleration type, Shock and Whip-like acceleration type, and Whip-like acceleration type \citep{yok96}. Therefore, the magnetic reconnection plays an important role in the transportation of filament material. The jet is a quite sufficient and direct way to carry dense and cool plasma from the lower atmosphere to upper atmosphere.

It is found that the apparent plane-of-sky velocity of the bright material in the SDO/AIA 304 $\rm \AA$ channel is about 160 km/s during the Jet A, while the Doppler velocities derived from the Si IV and H$\alpha$ line diagnostic is 10-20 km/s during the Jet B. For the order of magnitude difference in speed, it might be associated with the topology structure of the magnetic field of the filament. The plasma move along the magnetic field lines due to the low $\beta$ (the ratio of gas to magnetic pressures) in the corona and the magnetic field structure of the filament is almost horizontal (see Fig.\ref{figure9}). Furthermore, the filament is close to the center of the solar disk. Thus, it is reasonable that the plane-of-sky velocity of the bright material is significantly faster than the line of sight Doppler velocity from the Si IV and H$\alpha$ diagnostic.

Observationally, some coronal jets are associated with magnetic flux emergence or cancellation, which are believed as a source of significant mass and energy to input into the upper solar atmosphere \citep{shi98,wan99,liu04,iso07,ada14,rao16,hon17,li17}. However, there are also some studies of coronal jets, which are not associated with clear signatures of flux emergence \citep{moo13,moo15,ste15,ste16,ste17}. In many MHD simulations of coronal jets, the jets can be caused by the magnetic reconnection between an emerging magnetic flux and the pre-existing magnetic field \citep{yok95,yok96,mor08,mor13,che14}. The other signature during the jet is accompanying photospheric flux cancellation, which would be considered as a consequence of the magnetic reconnection after the jets. \cite{shi98} explained that the flux cancellation during the jet is due to the rate of flux emergence is smaller than the rate of photospheric reconnection. Sometimes, magnetic cancellation during jets is also due to the sink of inverse U-loops into the convection zone after the magnetic reconnection. Traditionally, jets are triggered by the magnetic reconnection between small closed magnetic field and adjacent open magnetic field \citep{mor08,che14,rao16}. In addition, some authors also reported that some jets are related to magnetic reconnection between two sets of closed magnetic fields, which are often associated with a fan-spine magnetic topology \citep{che15,wyp16,li17}. The hot or cool plasma are either spurted into outer atmosphere and become coronal mass ejection or fall back to low atmosphere, depending on the releasing energy of the magnetic field associated with jets \citep{yok96,can96}. In our study, we deduce that the magnetic emergence played an important role in producing these jets. Interestingly, the magnetic reconnection in our study was caused by closed and closed magnetic field instead of closed and open magnetic field. Therefore, the plasma jetted from jets could remain in the corona due to the magnetic dips, which usually existed in the closed magnetic field. On the other hand, we find that same dark threads appeared on the photosphere after one jet occurred at 18:02 UT on August 11. This is due to the increase of transverse magnetic field, which may be associated with the sink of small post-jet magnetic filed.

\begin{figure}[ht!]
\figurenum{11}
\plotone{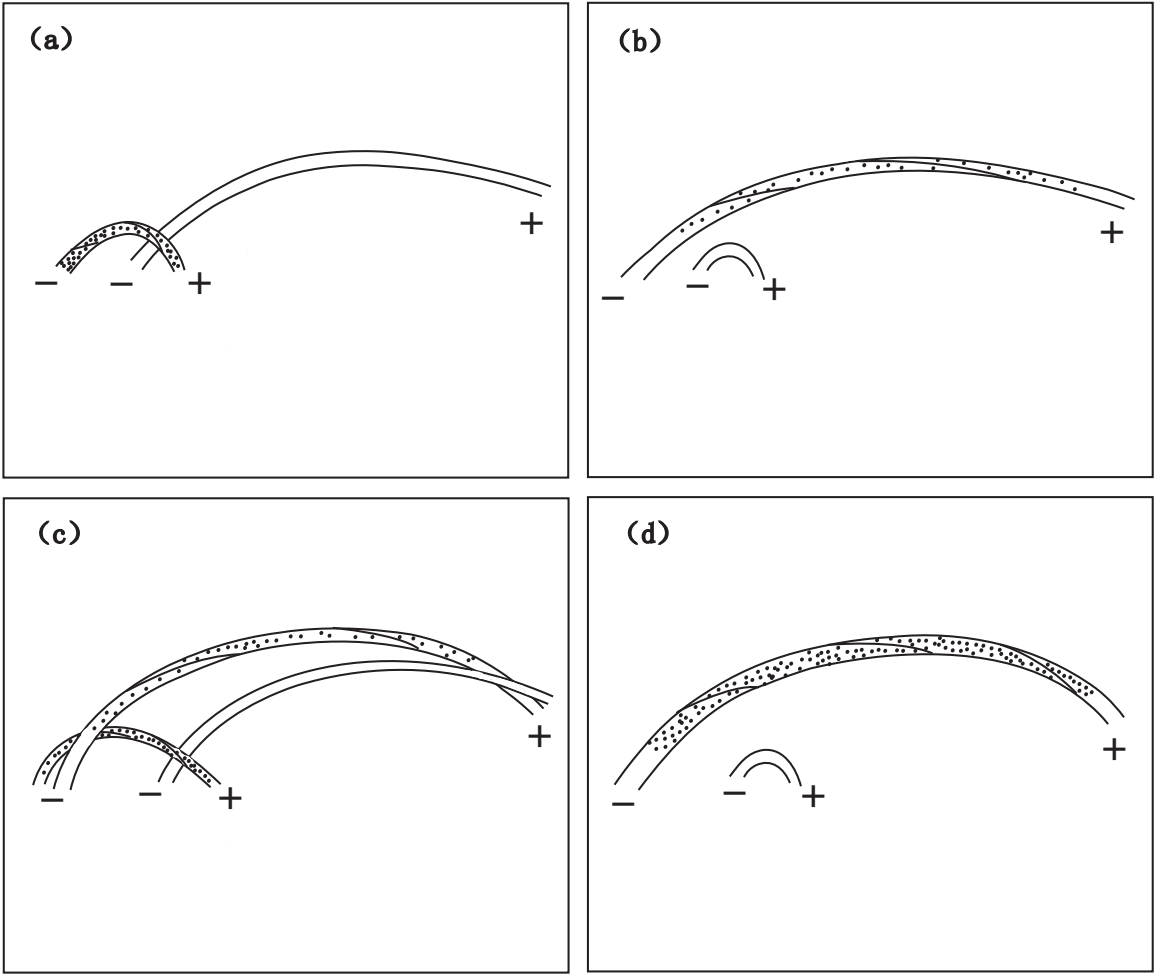}
\caption{Cartoon showing the formation of the filament by jets. \label{figure10}}
\end{figure}

The untwisting motion is often found in the jet events, which suggests that magnetic helicity stored in the closed emerging flux is transferred to the outer corona \citep{pik98,ste10,cur12,she11,sch13,moo15}. \cite{shi86} suggested the sudden release of the magnetic twist into an open flux tube is most likely to be due to the reconnection between a twisted loop and the open flux tube during the jet. Furthermore, this model was explained as the untwisting motion during the jet by other authors \citep{can96,moo15}. Similarly, we also find that the injected plasma was rotating during a jet. We deduce that the emerging closed magnetic field was a highly twisted structure. Due to magnetic reconnection between emerging closed magnetic field and closed magnetic field, the twist existing in the emerging closed magnetic field should be transformed to the closed magnetic field of the filament. Therefore, the jets not only injected the cool plasma into the filament height, but also influenced the configuration of the magnetic field of filament. The magnetic field topology of the filament became more and more non-potential after a series of jets, which is due to that the twist was transformed from the emerging magnetic field to the filament during periods of the jets. We draw some cartoons to illustrate this scenario. While the twisted emerging magnetic field reconnected with the pre-existing closed magnetic field, the twist was transformed to the long post-reconnected magnetic field and the material was injected into the long post-reconnected magnetic field (see Fig.\ref{figure10} (a)-(b)). After the magnetic reconnections occurred between emerging magnetic fields and the pre-existing closed magnetic fields for several times, the magnetic field became more and more twisting and the injected material was increasing (see Fig.\ref{figure10} (c)-(d)). The difference from the model suggested by \cite{shi86} is that the twist is transformed to the long post-reconnected magnetic field instead of releasing in the open flux tube. We conjecture that the jetted plasma fell down to other foot-point in early stage easier might be related to the local magnetic structure. At the beginning, less twisted structure existed in the filament magnetic structure. Therefore, the jetted plasma were hardly captured by the local magnetic field. As a series of jets happened and transformed the twist into the filament, the filament magnetic structure became more and more twisting and easily captured the jetted plasma. Thus, it is possible that material injection and magnetic structure of filament should form at the same time, especially for active-region filaments.

\acknowledgments
The authors thank the referee for constructive suggestions and comments that helped to improve this paper. SDO is a mission of NASA's Living With a Star Program and Hinode is a Japanese mission developed and launched by ISAS/JAXA, with NAOJ as domestic partner and NASA and STFC (UK) as international partners. It is operated by these agencies in co-operation with ESA and NSC (Norway). The authors are indebted to the SDO, GONG/NSO and Hinode teams for providing the data. This work is supported by the National Science Foundation of China (NSFC) under grant numbers 11603071,11503080, 11633008, 11527804, the Yunnan Talent Science Foundation of China, the Youth Innovation Promotion Association CAS under number 2011056, Yunnan Key Science Foundation of China under number Y8YJ061001, the grant associated with project of the Group for Innovation of Yunnan province, the Joint Research Fund in Astronomy (U1531140) under cooperative agreement between the National Natural Science Foundation of China (NSFC) and Chinese Academy of Sciences (CAS). The BBSO operation is supported by NJIT and US NSF AGS-1821294 grant. GST operation is partly supported by the Korea Astronomy and Space Science Institute and Seoul National University and by the strategic priority research program of CAS with Grant No. XDB09000000.


\end{document}